# Log-Periodic Oscillation Analysis and Possible Burst of the "Gold Bubble" in April – June 2011

***Tsirel, Sergey V.***, VNIMI, St. Petersburg, Russia
***Akaev, Askar***, "Complex System Analysis and Mathematical Modeling of the World Dynamics" Project, Russian Academy of Sciences
***Fomin, Alexey***, "Complex System Analysis and Mathematical Modeling of the World Dynamics" Project, Russian Academy of Sciences
**Korotayev, Andrey V.**, Russian State University for the Humanities, Moscow and "Complex System Analysis and Mathematical Modeling of the World Dynamics" Project, Russian Academy of Sciences

This working paper analyzes the gold price dynamics on the basis of methodology developed by Didier Sornette. Our calculations indicate that this dynamics is close to the one of the "bubbles" studied by Sornette and that the most probable timing of the "burst of the gold bubble" is April – June 2011. The obtained result has been additionally checked with two different methods. First of all, we have compared the pattern of changes of the forecasted timing of the gold bubble crash with the retrospective changes of forecasts of the oil bubble crash (that took place in July 2008). This comparison indicates that the period when the timing of the crash tended to change is close to the end, and the burst of the gold bubble is the most probable in May or June 2011. Secondly, we used the estimates of critical time for the hyperbolic trend (that has been shown in our previous publications to be typical for many socioeconomic processes). Our calculations with this method also indicate May – June 2011 as the most probable time of the burst of the gold bubble. Naturally, this forecast should not be regarded as an exact prediction as this implies the stability of the finance policies of the USA, European Union, and China, whereas a significant intervention of giant players (like the Federal Reserve System, or the Central Bank of China) could affect the course of the exchange game in a rather significant way. We also analyze possible consequences of the burst of the "gold bubble". On the one hand, it is obvious that such a collapse leads to huge losses or even bankruptcies of many of the major participants of exchange games and their dependent firms and banks. Therefore, the immediate market reaction will be entirely negative. Negative impact on the market will be amplified by numerous publications in the media and business press, drawing analogies with the events of the early 1980s and earlier similar events, as well as by losses of the shareholders of bankrupt companies. The current instability of major world currencies and the most powerful world economies, unfortunately, does not preclude the escalation of a short-term downswing into the second wave of economic and financial crisis. On the other hand, investments in gold, which caused the very "gold rush", also lead to the diversion of funds from stock market investments and to the reduction in the production of goods and services. If at the time of the collapse some promising areas of investment appear in the developed and / or developing countries, the investment can move to those markets, which, on the contrary, could contribute to the production of new goods and services, accelerating the way out of the crisis.

In a number of seminal works by Didier Sornette, Anders Johansen and their colleagues (Sornette, Sammis 1995; Sornette, Johansen 1997, 2001; Johansen, Sornette 1999, 2001; Johansen *et al.* 1996; Sornette 2004; etc.) it has been demonstrated that accelerating log-periodic oscillations superimposed over an explosive growth trend that is described with a power-law function with a singularity (or quasi-singularity) in a finite moment of time $t_c$, are observed in situations leading to crashes and catastrophes. They can be analyzed because their precursors allow the forecasting of such events. One can mention such examples as the log-periodic oscillations of the Dow Jones Industrial Average (DJIA) that preceded the crash of 1929 (*e.g.*, Sornette, Johansen 1997), or the changes in the ion concentrations in the underground waters that preceded the catastrophic Kobe earthquake in Japan on the 17$^{th}$ of January, 1995 (*e.g.*, Johansen *et al.* 1996), which are also described mathematically rather well with log-periodic fluctuations superimposed over a power-law growth trend.

$$* \quad * \quad *$$

After the gold standard policy (see, *e.g.*, Eichengreen, Flandreau 1997) had come to its end, gold began to be transformed into a "last reserve anchor", a commodity in which the players tend to invest very intensively in such contexts when one observes the running out of other markets and commodities, and the investment into which one could preserve unsafe money. After the end of the gold standard era the first "gold bubble" formed in 1979–1980 during the second jump of the oil prices and was a symptom of prolonged economic crisis (see Figs 1 and 2).

During the current global crisis the gold prices grew in 2008 significantly less than the prices of oil and other raw materials, as if indicating that currently gold is the main reserve currency. However, oil, gas, coal, and metals are not means of hoarding only; they are also very important raw materials for various industries; that is why the jump in their prices brought the global crisis nearer and accelerated it. Now, *post factum*, it is almost evident that the surge in raw materials' prices is simultaneously an indicator of the closeness of crisis and a factor that brings it closer, but such an apparently self-evident idea only became obvious during the crisis itself.

In the context of the crisis, gold started to restore the role of the reserve asset, which accelerated the growth of the gold prices. The countercrisis policy conducted by the leading Western governments influenced the growth of gold prices even more. As is known, it was based on large infusions of public funds into the economy, and, naturally, led to the acceleration in the process of gold prices increase that had begun before the crisis. The weakening of the existing world currencies and the reluctance of China to convert yuan into a world currency in some way returned to gold a part of its former functions. In the absence of a fixed price for gold (in other words, floating gold content of the major currencies), this leads to the formation of a bubble, rather similar to bubbles in commodity markets or real estate. It is quite obvious that this is no ordinary bubble, and the gold market or, conversely, the gold content of the major currencies, is not exactly a free market game, and even those laws that characterize the market game to some extent suit this case only conditionally. A Congress bill or simply Bernanke's decision seems to be sufficient to significantly affect the process.

**Fig. 1.**    Yearly gold price dynamics, 1970–2010

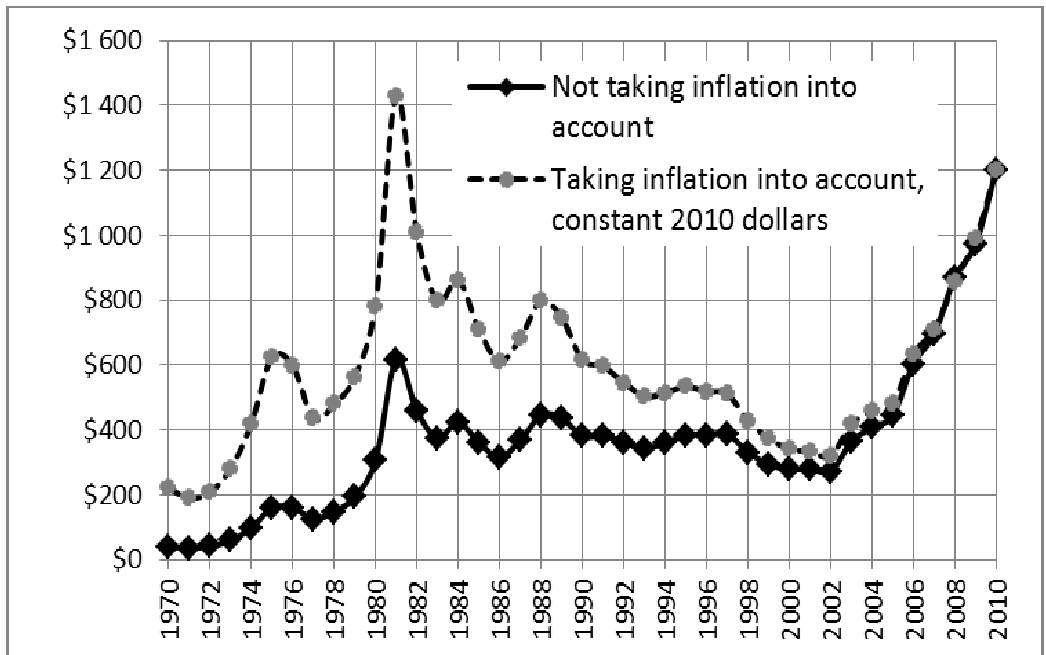

*Note:* yearly London fixing averages per a troy ounce of gold. *Sources*: World Gold Council database. URL: http://www.research.gold.org/prices/ (gold prices for 1970–2009); USA Gold Reference Library database. URL: http://www.usagold.com/reference/prices/history.html (average price for January 4 – November 12, 2010); World Development Indicators Online (Washington, DC: World Bank, 2010), URL: http://data.worldbank.org/data-catalog/world-development-indicators (data on USA inflation).

**Fig. 2.**    Yearly oil price dynamics, 1970–2010

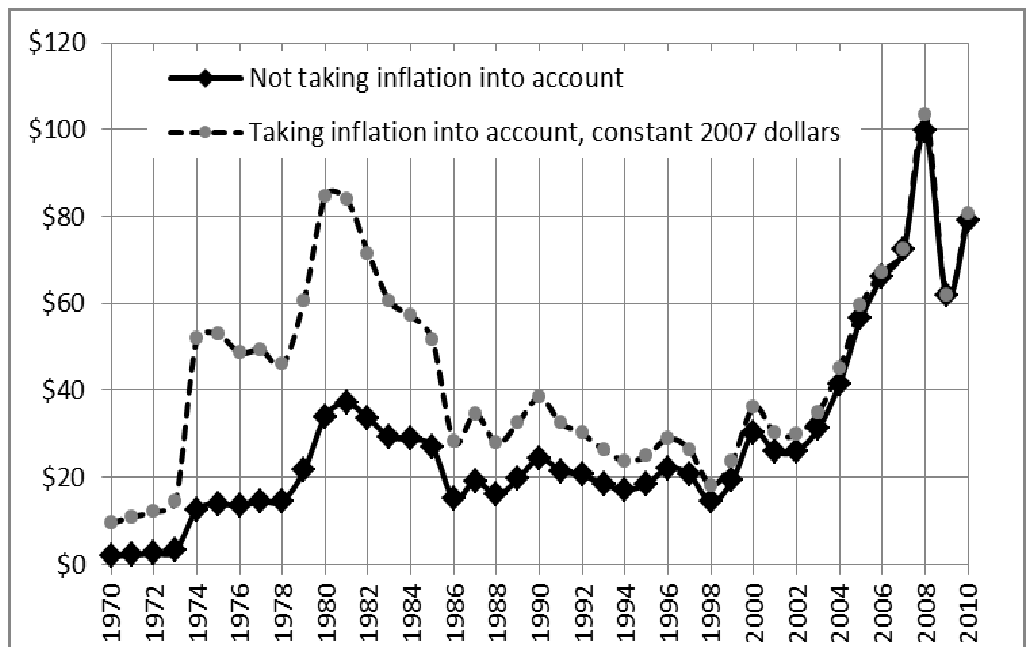

*Note:* 1970–1973 prices are the official price of a barrel of Saudi Light, 1974–1985 prices are refiner acquisition costs of imported crude oil, 1986–2010 prices are spot prices for West Texas Intermediate at Cushing, OK. *Sources:*

Earth Policy Institute (Washington, DC, 2010) database (URL: www.earth-policy.org/datacenter/xls/update67_5.xls, oil prices for 1970–2006); U. S. Energy Information Administration database. URL: http://www.eia.doe.gov/dnav/pet/ pet_pri_spt_s1_a.htm (цены на нефть за 2007–2010 годы); World Development Indicators Online (Washington, DC: World Bank, 2010), URL: http://data.worldbank.org/data-catalog/world-development-indicators (data on the inflation in the U.S.).

Nevertheless, we still make an assumption that in the first place we are not dealing with the complex relationships of the Central Banks of the great powers, but with a "market bubble", which, as Sornette and others have shown (Sornette, Johansen 1997, 1998, 2001; Johansen, Sornette 1999, 2001; Johansen, Sornette, Ledoit 1999; Sornette 2004, etc.), is characterized by "log-periodic" price fluctuations. We also assume that central bank policy of great powers in the critical period of inflation and the collapse of the gold bubble will remain fairly stable. We use Sornette's approach to try to estimate when the "gold bubble" could burst.

The basic equation derived by Sornette and tested on many historical examples of bubbles has the following form:

$$\ln[p(t)] = A - m\,(t_c - t)^\alpha \{\,1 + C\cos[\omega\ln(t_c - t) + \varphi]\,\}, \tag{1}$$

or

$$p(t) = A - m\,(t_c - t)^\alpha \{\,1 + C\cos[\omega\ln(t_c - t) + \varphi]\,\}, \tag{1a}$$

where $p(t)$ is gold price at the moment $t$ (further on we operate with daily London gold fixings by PPI index to the 1982 dollar); $t_c$ is the "critical time"; $A$, $m$, $C$, $\alpha$, $\omega$, and $\varphi$ are constants which are to be defined on the basis of data on gold prices from the start of the bubble formation till the forecast moment. Sornette (2004) recommends equation (1) for the analysis of the long-term development of bubbles, whereas he suggests that both equations can be used for a shorter-term (less than two years) development of them.

In these equations $m\,(t_c - t)^\alpha$ describes the main trend of the growth dynamics; with the approaching of the critical point $t_c$, price $p(t)$ approaches the maximum value $A$. Against this background periodic oscillations with reduced period take place. These oscillations (Sornette calls them log-periodic oscillations) are described by the second member $C\,(t_c - t)^\alpha \cos[\omega\ln(t_c - t) + \varphi]$ multiplied by the first considered member (whose value decreases with the lapse of time). Thus, the oscillations amplitude is steadily declining.

Of course, not only infinite, but also very high frequency of the market price fluctuations is really impossible. Achieving the oscillation frequency of high values means an increased risk of crisis. On average, in the examples treated by Sornette (2004), the crisis occurs 1.4 months before the critical point $t_c$.

At the moment of crisis the growth of $p(t)$ stops and there frequently begins its sharp decline (market crash). The second option marked by Sornette is a scenario when the bubble softly "blows off", the price starts to decline more or less in reverse order with respect to the sequence in which it grew up ("anti-bubble"). This scenario is the least damaging to the economic and socio-political perspective and in order to direct the process to a smoother path various state regulatory mechanisms are used.

In the practical implementation of the described methods of forecasting there rises the question of choosing the time interval for which the parameterization will be implemented. In Sornette's monograph (2004) this issue is solved empirically, *i.e.* such an interval is chosen within which both the main growth trend and the oscillations with reduced periods described by equation (1) are clearly visible.

Analysis of the growth dynamics of the gold price from the beginning of 1973 to December, 2010 showed that the time interval that should be used to forecast the date of the collapse of the gold market lies roughly after 2002.

Growth dynamics of the nominal price of gold is shown in Fig. 3:

**Fig. 3.**      Daily gold price dynamics, 1973–2010, US dollars

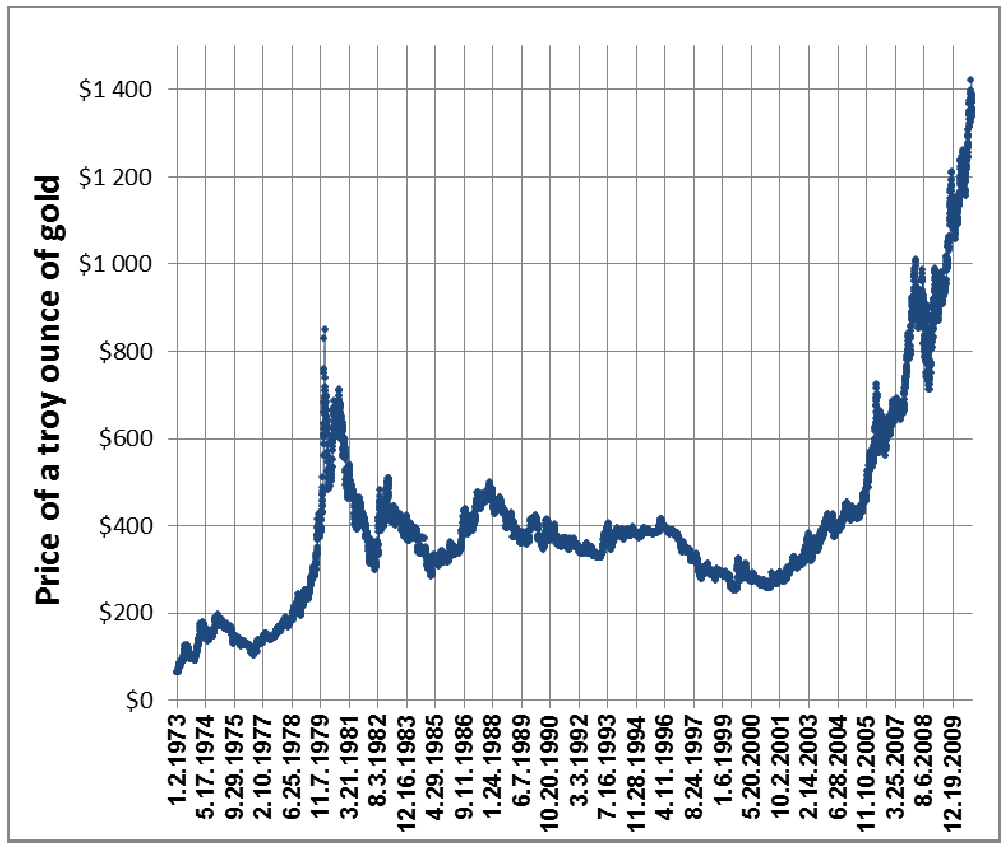

As shown in Fig. 3, rapid price growth began approximately after 2002 against the background of relatively small oscillations. Namely this growth, as Sornette was the first to note, is well described by equation (1).

In the considered scale after 2002 the oscillations are seen not very well. In a larger scale, oscillations in the real price have the form shown in Fig. 4:

**Fig. 4.** Log-periodic oscillations in the gold price dynamics,
June 11, 2003 – December 2, 2010 (taking inflation into account;
constant 1982 dollars)

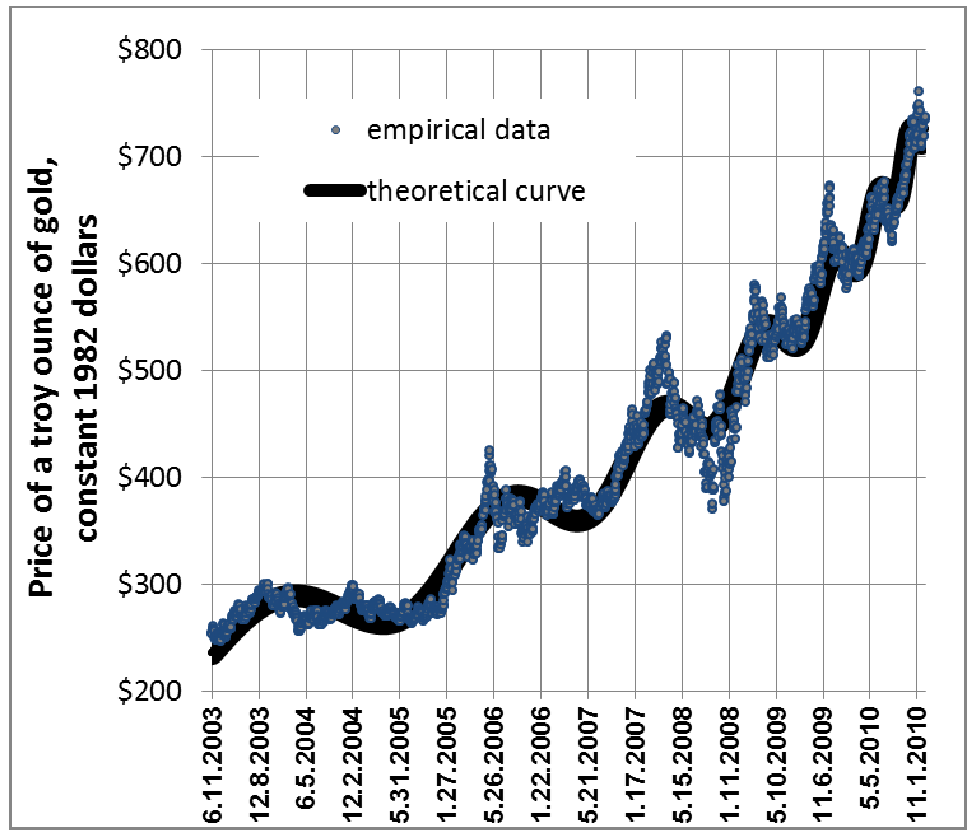

In Fig. 4, grey markers indicate daily gold price between June 11, 2003 and December 2, 2010, whereas the smooth thick black line has been generated by equation (1a), with parameters chosen by the least squares (see equation (2) below).

In accordance with Sornette's (2004) recommendations we tried to use equation (1); however, our calculations have indicated that sum of deviation squares for the price logarithm has no expressed minimums in the area of the real values of parameters. That is why equation (1a) has been chosen.

As our calculations have shown, the critical/singular point can be identified here as $t_c$ = 2011.45 (*i.e.* June 14, 2010). The smooth thick line in Fig. 4 has been generated by the following equation:

$$p(t) = 1220.41 - 570.35\,(2011.45 - t)^{.267}\,\{\,1 + 0.036\cos[15.86\ln(2011.45 - t) - 34.8]\,\}. \quad (2)$$

Though, contrary to Sornette's recommendations, we have chosen an equation without a logarithm in its left-hand side, the values of parameters of equation (2) mainly lie in the range of characteristic values calculated in Sornette's monograph (2004). For example, in this book the parameter $\omega$ (= 15.86 in our case) in different cases varies in the range between 2.9 and 15. This parameter is related to the so-called scaling factor $\exp(2\pi/\omega) = 1.487$, which shows by how many times the duration of oscillation is reduced from period to period. Exponent (.267) is close to its characteristic value .3, according to Sornette's calculations. In general, the calculations described above fit well in a series of calculations performed by the method's author.

Basing on Sornette's observation that collapses tend to occur *c.*1.4 months before the moment of singularity calculated according to his method, **one would forecast that the collapse in gold prices (or the termination of their growth) is most likely to occur in late April – early May 2011.**

Two approaches have been applied in order to check our result.

The first approach is based on the analogy with the oil bubble (see Sornette, Woodard, and Zhou 2009). Though the log-periodic oscillations are visible in the oil price dynamics (see Fig. 5) much less distinctly than in the gold price dynamics, equation (1a) turns out to detect the critical time in a rather accurate way. The calculated critical moment in this case has turned to be August 31, 2008 (note that we used data points up till April 25, 2008 in those calculations). According to to Sornette (2004), the actual start of the collapse of a respective indicator tends to start *c*1.4 months before the critical moment, which in our case corresponds to *c*July 18, 2008. Note that the actual crash of the oil prices started on July 14, 2008.

**Fig. 5.** Log-periodic oscillations in the oil price dynamics, January 18, 2007 – April 25, 2008 (taking inflation into account; constant 1982 dollars)

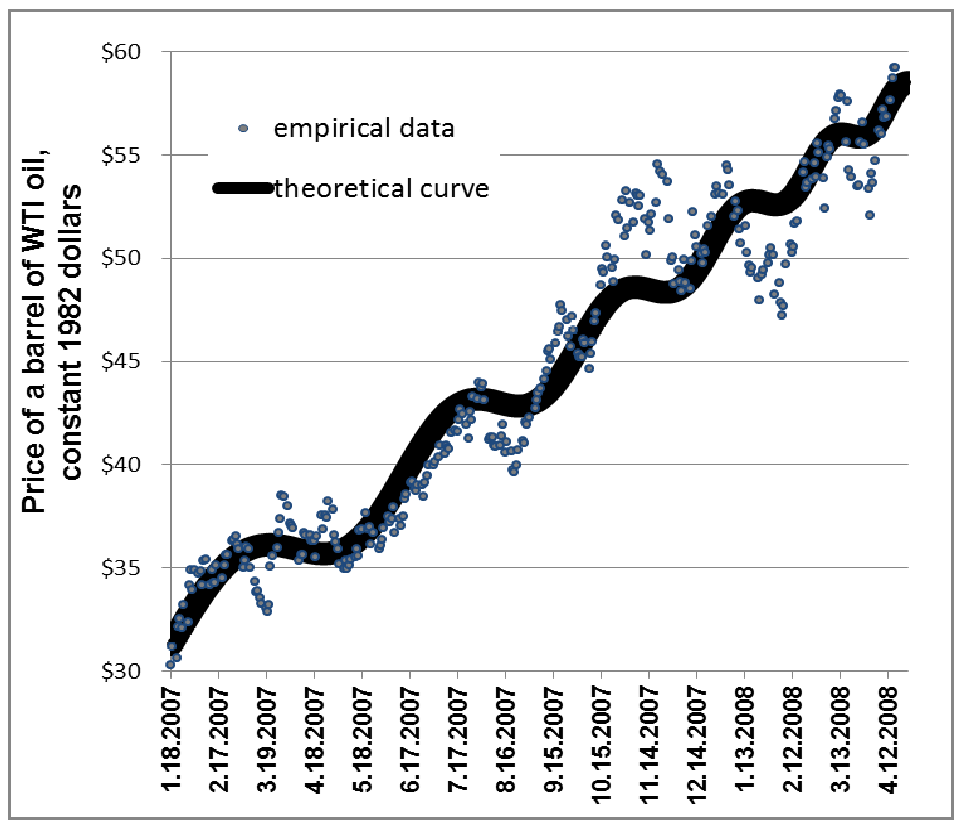

In Fig. 5, grey markers indicate daily gold price between January 18, 2007 and April 25, 2008, whereas the smooth thick black line has been generated by equation (1a), with parameters chosen by the least squares (see equation (3) below):

$$p(t) = 66.9 - 22.84\,(2008.67 - t)^{.915}\,\{\,1 + 0.056\,\cos[22.1\,\ln(2008.67 - t) + .47]\,\}. \tag{3}$$

We have investigated the following question: starting from which time point is a sufficiently precise forecast possible? In order to do this we have made calculations for the "oil bubble", using various time points of the termination of observation. This way we have actually imitated attempts to forecast the burst of the 2008 oil bubble on, say, May 15, 2008, or March 10, 2008. As in the case of the "gold bubble" we have used for our calculations equation (1a) rather than equation (1), because in this case equation (1) has not revealed local minimums either in the area of our interest. In general, the question of the area of applicability of equation (1) seems to be still open.

Our calculations indicate (Fig. 6) that with a later time of the end of observations the temporal distance between the critical moment and the time of the end of observation first increases. However, approximately two months before the crash of the oil prices (the red line) the position of the critical point stabilizes, and the interval between the critical point and the actual bubble burst turns out to be about 1 – 1.5 months.

**Fig. 6.** The dependence of the calculated position of the critical point on the moment of the discontinuation of the observation of the oil prices

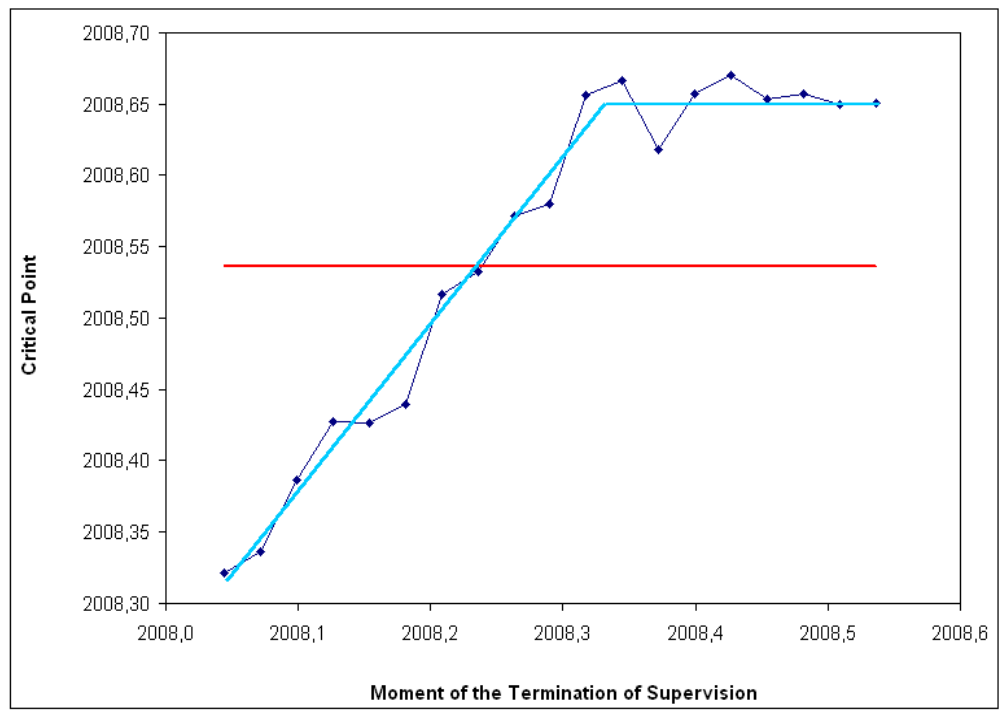

Basing on the dependence obtained, we have tried to make a similar curve for the gold prices.

The curve for the gold prices has turned out to be not exactly the same as the one for the oil prices. In contrast to the oil bubble in Fig. 7 we do not see a clear stabilization of the forecasted critical point date, but the discontinuation of strong oscillation and a gradual approach to the (approximately) finite value during 2010. The discontinuation of strong oscillations is also indicated by the cessation of sharp changes in parameter A which characterizes the gold price at the critical point (Fig. 8).

**Fig. 7.** The dependence of the calculated position of the critical point (Y-axis) on the moment of discontinuation of the observation of the gold prices (X-axis)

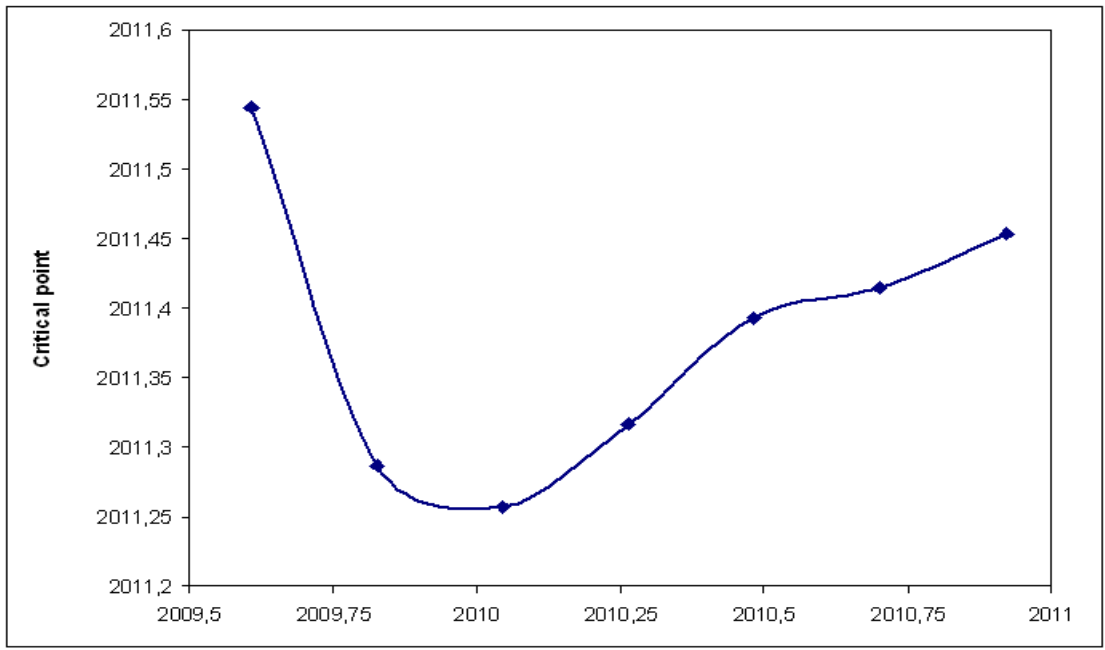

**Fig. 8.** The dependence of the calculated position of parameter A (Y-axis) on the moment of discontinuing the observation of the oil prices (X-axis)

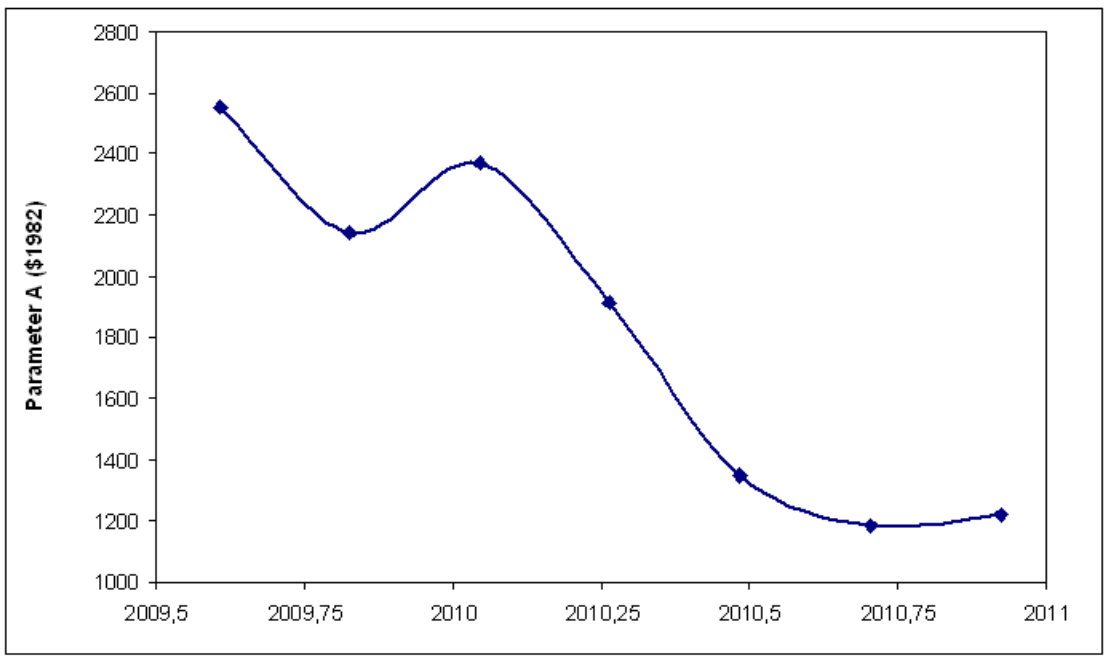

Various methods of extrapolating the curve shown in Fig. 7 indicate that the finite position of the critical point presumably lies in the range between June 20$^{th}$ and July 12$^{th}$, 2011. Thus, taking into account the average length of the time interval between the critical point and the real date of the price growth cessation/collapse, the most probable time of the gold bubble burst is May, 2011.

The calculations also allow approximately defining the gold price at the growth cessation (start of collapse) point. The estimates of A = $1150–1250 (in constant 1982 dollars) shown in Fig. 8 approximately correspond to the gold price at the most possible moment of growth cessation/start of collapse (1.4 months before the singularity point). Note that this corresponds to $1500–1700 range in current prices.

The second method of checking is based on another approximation of the price growth curve. An attempt has been made to forecast the point of collapse using log-periodic oscillations superimposed on a hyperbolic trend typical for many social processes (population growth before the demographic transition, global GDP growth and energy production per capita, *etc.*) (see, *e.g.*, von Foerster, Mora, Amiot 1962; Johansen, Sornette 1999; Tsirel 2004; Khaltourina, Korotayev, Malkov 2006; Malkov, Korotayev, Khaltourina 2006; Korotayev, Malkov, Khaltourina 2006; Markov, Korotayev 2007; Korotayev 2005, 2006, 2007, 2009, *etc.*). Such a method of calculating the point of collapse is deprived of many advantages, which Sornette's method possesses: a simultaneous detection of the crash moment through the trend and the oscillation periods, the lack of a significant influence of the weight of events (recent observations at the highest price play a greater role for the trend, while the first observations are more important for the periods, as the periods were longest then and each period contained a lot of data points), and the possibility to take into account the oscillation amplitude.

However, the unification of the estimation of the collapse point position through the trend analysis and through the analysis of the oscillatory process proposed by Sornette is based on rather rigid assumptions, because the effects that they experience are not identical – the trend strongly depends on more general and long-term tendencies, while variations depend on the current events. Therefore, the estimates free of such a rigid connection possess some independent value.

The analysis of many processes of rapid price increases in recent years (for example, the process of oil prices' growth in 2000–2008) has demonstrated that the amplitude of respective log-periodic oscillations had no pronounced trends. Therefore, we have used the following equation:

$$p(t) = A\,(t_{c1} - t)^B + C \cos[\omega \ln(t_{c2} - t) + \varphi]. \qquad (4)$$

In this equation (unlike in equation (1)) the amplitude does not change.

The first step was to describe the hyperbolic trend. It should be noted that the trend equation is highly dependent on the starting point of the hyperbolic growth (though log-periodic deviation from the trend is much less dependent on the starting point). Also, as in Sornette's method, the selection of the starting point was made "by eye", referring to the analyzed curve. For the selected starting point in August 2002 we have arrived at a hyperbola of the following kind (Fig. 9).

$$p(t) = A / (t - 2015)^{1-1.05}. \qquad (5)$$

In this equation gold prices would theoretically go to the infinity (singularity point 1) approximately at the end of December 2014 (or autumn 2014 with an earlier starting point of hyperbolic growth trajectory).

The second step was the approximation of the deviations from the trend in the form of log-periodic oscillations. First, each point (the same as in Sornette's method) was given the same weight. Herewith, the end of this process (singularity point 2) fell on July 6, 2011 (Figs 10 and 11), which finds itself into the above mentioned range. Basing on this estimation, the most probable time of the gold bubble burst is May – June 2011, and the maximum gold price calculated according to (5) will reach the level of $1500–1600 in current prices.

**Fig. 9.** Hyperbolic trend in the gold price dynamics, 2002–2010

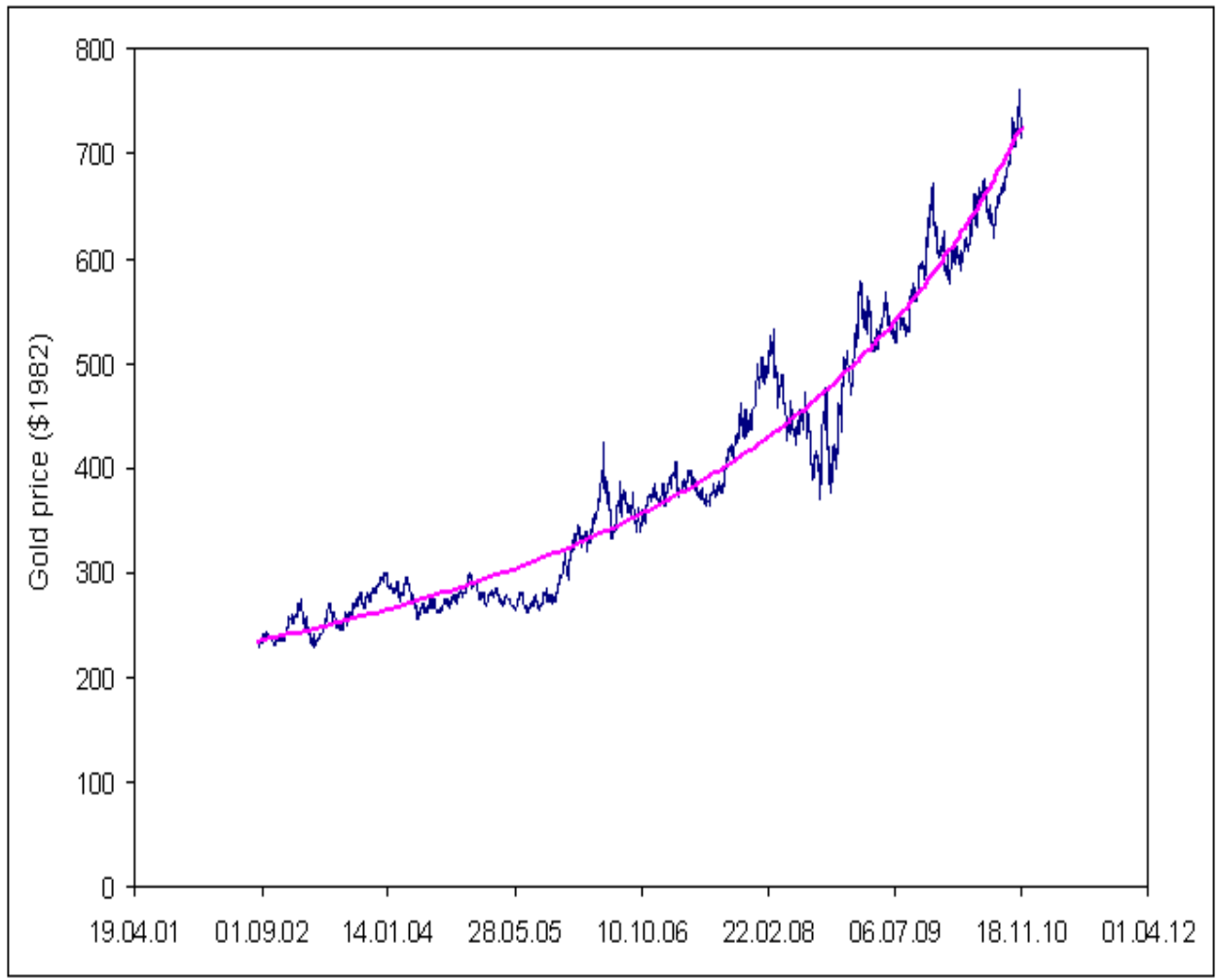

Therefore, the most likely estimate of the period when the "gold bubble" is likely to burst is May – June 2011.

**Fig. 10.** Deviations from hyperbolic trend in the gold price dynamics, 2002–2010

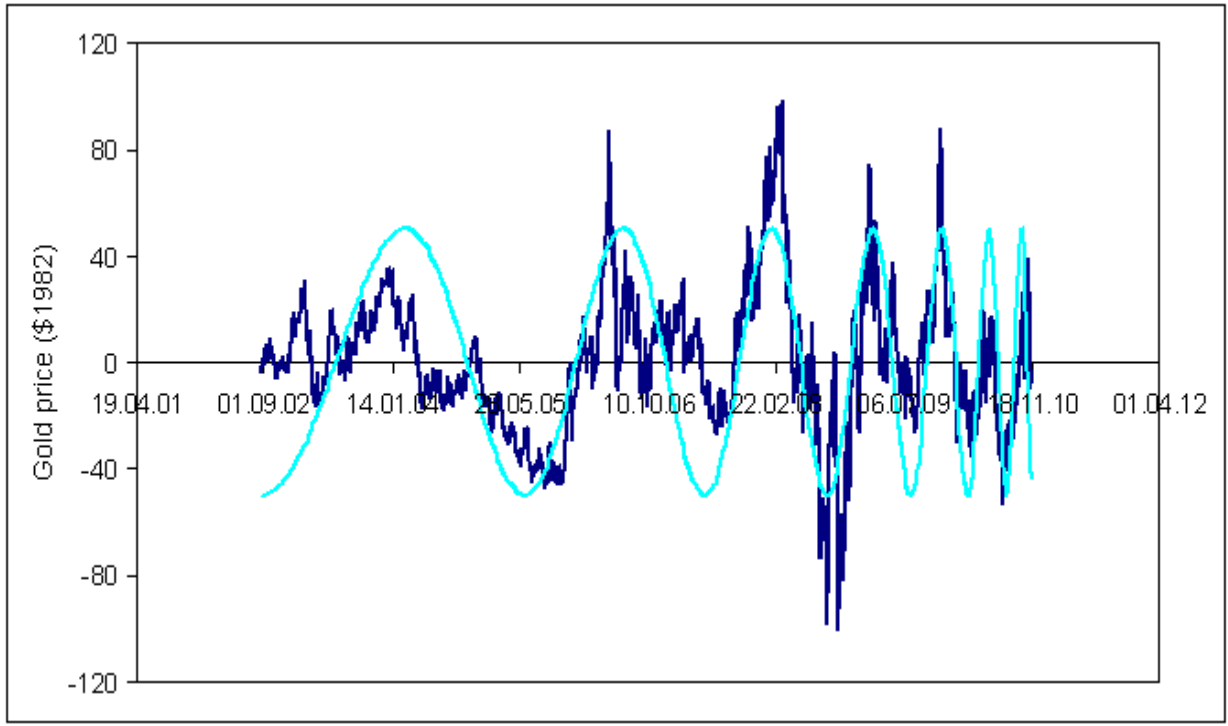

**Fig. 11.** Logarithmic scale deviations from hyperbolic trend in the gold price dynamics, 2002–2010 (X-axis indicates temporal distance in days from the critical time point)

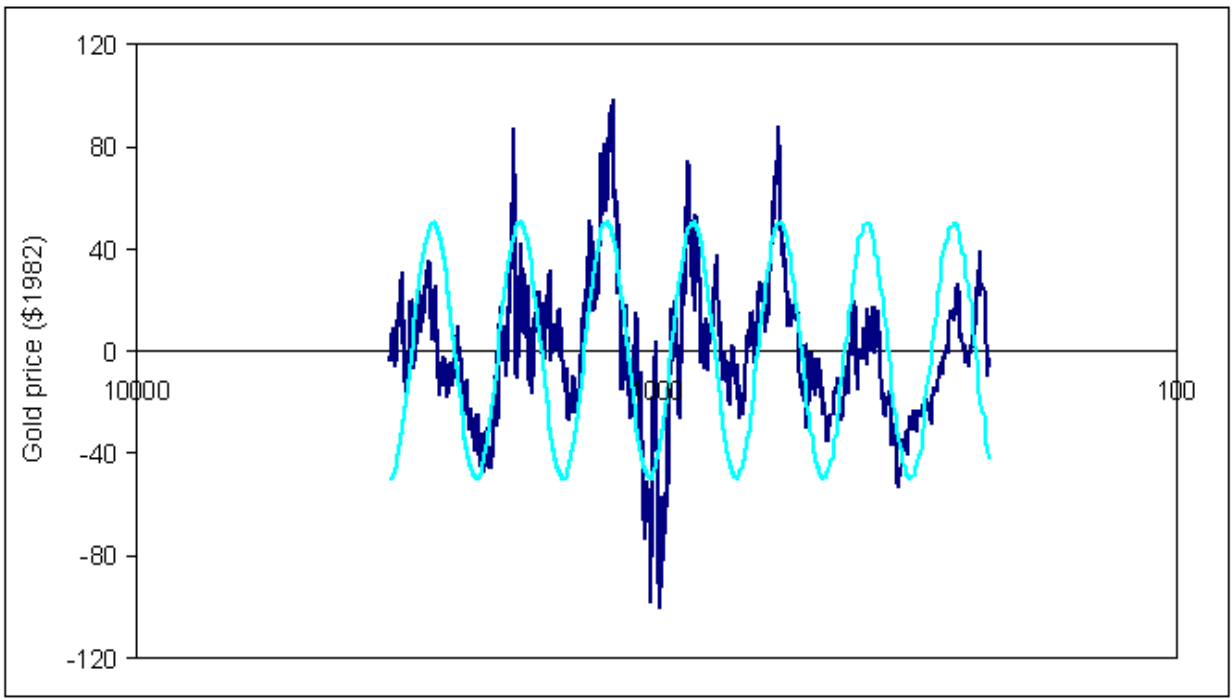

At the same time, as stated in the beginning of this article, all these calculations should not be given the meaning of an accurate prediction of the events. The calculations performed by Sornette's method and other estimates are based on the idea that all players will try to maximize profit just in this market, and will not believe in forecasts like the one we propose (otherwise we shall deal with a "non-self-fulfilling prophecy"). If a very large player, such as the Central Bank of China, intervenes with other goals (longer-term, depending on political calculations, etc.), then the foundation upon which all such forecasts are based disappears.

Will the probable collapse of the "gold rush" signify a new wave of crisis?

On the one hand, it is obvious that such a collapse leads to huge losses or even bankruptcies of many of the major participants of exchange games and their dependent firms and banks. Therefore, the immediate market reaction will be entirely negative. Negative impact on the market will be amplified by numerous publications in the media and business press, drawing analogies with the events of the early 1980s and earlier similar events, as well as by losses of the shareholders of bankrupt companies. The current instability of major world currencies and the most powerful world economies, unfortunately, does not preclude the escalation of a short-term downswing into the second wave of economic and financial crisis.

On the other hand, investments in gold, which caused the very "gold rush", also lead to the diversion of funds from stock market investments and to the reduction in the production of goods and services. If at the time of the collapse some promising areas of investment appear in the developed and / or developing countries, the investment can move to those markets, which, on the contrary, will contribute to the production of new goods and services and accelerate the way out of the crisis.

Rather than relying on the most optimistic scenario presented in the previous paragraph, let us suppose that the burst of "the golden bubble" will be followed by a short-term downswing, and the further developments will depend on the amount of perspective directions of investment at the moment, as well as on the reaction of the U.S. Federal Reserve and the Central Bank of China to the events. In absence of sharp uncoordinated actions there could be certain grounds to expect the reduction of the influence of gold prices on world markets and the continuation trend of slow and unstable recovery.

### Note

A short version of this text has already been published as a separate article (Akayev *et al*. 2010). However, that publication has not included the most technical part of this text (see pages 3–9 above) that we believe is the most interesting from the scientific point of view; so, we have decided (and have been supported by Didier Sornette at this point) that the whole version of this text also deserves the publication.

### Acknowledgement

We would like to express our gratitude to Julia Zinkina (Institute for African Studies, Russian Academy of Sciences) for her invaluable contribution to the preparation of the English version of this text.